\documentclass[lettersize,journal]{IEEEtran}
\usepackage{amsmath,amsfonts}
\usepackage{algorithmic}
\usepackage{algorithm}
\usepackage{array}
\usepackage[caption=false,font=normalsize,labelfont=sf,textfont=sf]{subfig}
\usepackage{textcomp}
\usepackage{stfloats}
\usepackage{url}
\usepackage{verbatim}
\usepackage{graphicx}
\usepackage{cite}

\begin{document}

\title{Multimode optical fiber beam-by-beam cleanup}

\author{Mario Ferraro,~\IEEEmembership{Member,~Optica,} Fabio Mangini, Yann Leventoux, Alessandro Tonello, Mario Zitelli~\IEEEmembership{Member,~Optica,} Yifan Sun,~\IEEEmembership{Member,~Optica,} Sebastien Fevrier, Katarzyna Krupa, Denis Kharenko, Stefan Wabnitz,~\IEEEmembership{Fellow Member,~Optica,} and Vincent Couderc






\thanks{
This work was supported by the European Research Council (ERC) under the EU HORIZON2020 Research and Innovation Program (740355), the Italian Ministry of University and Research (R18SPB8227), by Sapienza University of Rome (RG12117A84DA7437), the EU HORIZON2020 Marie Skłodowska-Curie program (713694), and the French research national agency (ANR-18-CE080016-01, ANR-10-LABX-0074-01).
(Corresponding author: Mario Ferraro.)}
\thanks{Mario Ferraro, Fabio Mangini, Mario Zitelli, and Yifan Sun are with the Department of Information Engineering, Electronics and Telecommunications (DIET),
Sapienza University of Rome, 00184 Rome, Italy (e-mail: mario.ferraro@uniroma1.it; fabio.mangini@uniroma1.it; mario.zitelli@
uniroma1.it; yifan.sun@uniroma1.it).}

\thanks{Stefan Wabnitz is with the Department of Information Engineering, Electronics and Telecommunications (DIET),
Sapienza University of Rome, 00184 Rome, Italy, and with CNR-INO, Istituto Nazionale di Ottica, Via Campi Flegrei 34, 80078 Pozzuoli (NA), Italy (e-mail: stefan.wabnitz@uniroma1.it).}

\thanks{Yann Leventoux, Sebastien Fevrier, Alessandro Tonello, and Vincent Couderc are with the XLIM Institute (UMR CNRS 7252), University of Limoges, 87060 Limoges, France (email: yann.leventoux@unilim.fr; sebastien.fevrier@unilim.fr; alessandro.tonello@unilim.fr; vincent.couderc@xlim.fr).}
\thanks{Katarzyna Krupa is with the Institute of Physical Chemistry, Polish Academy of Sciences, Warsaw, Poland (email: kkrupa@ichf.edu.pl).}

\thanks{Denis Kharenko is with the Institute of Automation and Electrometry SB RAS, 1 ac. Koptyug ave., Novosibirsk 630090, Russia, and with Novosibirsk State University, Novosibirsk 630090, Russia}

\thanks{Mario Ferraro, Fabio Mangini and Yann Leventoux contributed equally to this work.}
%
}



\maketitle

\begin{abstract}
We introduce and experimentally demonstrate the concept of all-optical beam switching in graded-index multimode optical fibers. Nonlinear coupling between orthogonally polarized seed and signal beams permits to control the spatial beam quality at the fiber output. Remarkably, we show that even a weak few-mode control beam may substantially enhance the quality of an intense, highly multimode signal beam. We propose a simple geometrical representation of the beam switching operation, whose validity is quantitatively confirmed by the experimental mode decomposition of the output beam. All-optical switching of multimode beams may find important applications in high-power beam delivery and fiber lasers.
\end{abstract}


\section{Introduction}
\IEEEPARstart{T}{he} future of optical communications relies on the capacity to deliver and process complex ultrafast optical signals. 
In the realm of optical beam delivery, multimode fibers (MMFs) are well suited for the transmission of high-power, structured optical laser beams. As a matter of fact, MMFs support the propagation of optical beams with complex temporal and spatial properties \cite{krupa2019multimode, picozzi2015nonlinear}. In addition, opening up the spatial degree of freedom is the key for upgrading the capacity of optical communication systems via the space-division-multiplexing technique \cite{richardson2013space}.
Another noteworthy advantage of MMFs with respect to their singlemode counterparts is their capability to support the propagation of both high-energy optical pulses and high-power beams. In this framework, MMFs naturally find applications for the power or energy up-scaling of fiber-based light sources, ranging from mode-locked ultrafast lasers \cite{wright2017spatiotemporal} to supercontinuum generators \cite{poletti2009dynamics, krupa2016spatiotemporal, lopez2016visible, eftekhar2017versatile}.
In addition, high peak power beam propagation in MMF allows for exploring intriguing fundamental effects and nonlinear phenomena, such as multimode solitons \cite{crosignani1981soliton,renninger2013optical, wright2015spatiotemporal, zitelli2021conditions}. 

Material nonlinearity plays a key role in all-optical signal processing devices: beyond birefringence, exploiting nonlinearity is the only way that two optical beams can interact. As a matter of fact, two beams propagating in a linear medium can only produce interference, but one can never change the properties of a beam by acting on another beam of light. Hence, nonlinear beam propagation has paved the way for signal processing based on all-optical technologies. This basic concept has been largely explored in the recent past \cite{wabnitz2015all}. Many studies of nonlinear all-optical signal processing devices have been reported in the literature, including, e.g., logic gates, switches, amplifier and transistors. 
All-optical signal processing have been proposed in several platforms, ranging from gases \cite{dawes2005all} to semiconductors \cite{almeida2004all, ballarini2013all}, glasses \cite{asobe1997nonlinear, digonnet1997resonantly}, quantum dot cavities \cite{volz2012ultrafast} and integrated optical circuits \cite{nozaki2010sub}.





In this work, we introduce a new optical platform for nonlinear optical signal processing in the form of multimode guided beams. 
Specifically, we demonstrate the all-optical switching of a multimode light beam, controlled by a different, copropagating multimode beam. The physical mechanisms for switching is the intensity dependent refractive index, or Kerr effect, in a standard graded-index (GRIN) MMF. When considering the propagation of a single multimode light beam in such a fiber, the same Kerr effect may self-induce a significant improvement of its spatial quality at the fiber output. Such a nonlinear phenomenon has been firstly observed in 2016 by Krupa et al. \cite{krupa2016observation}, and it is known as spatial beam self-cleaning \cite{liu2016kerr, krupa2017spatial}. \IEEEpubidadjcol 

\begin{figure*}[ht!]
\centering\includegraphics[width=15cm]{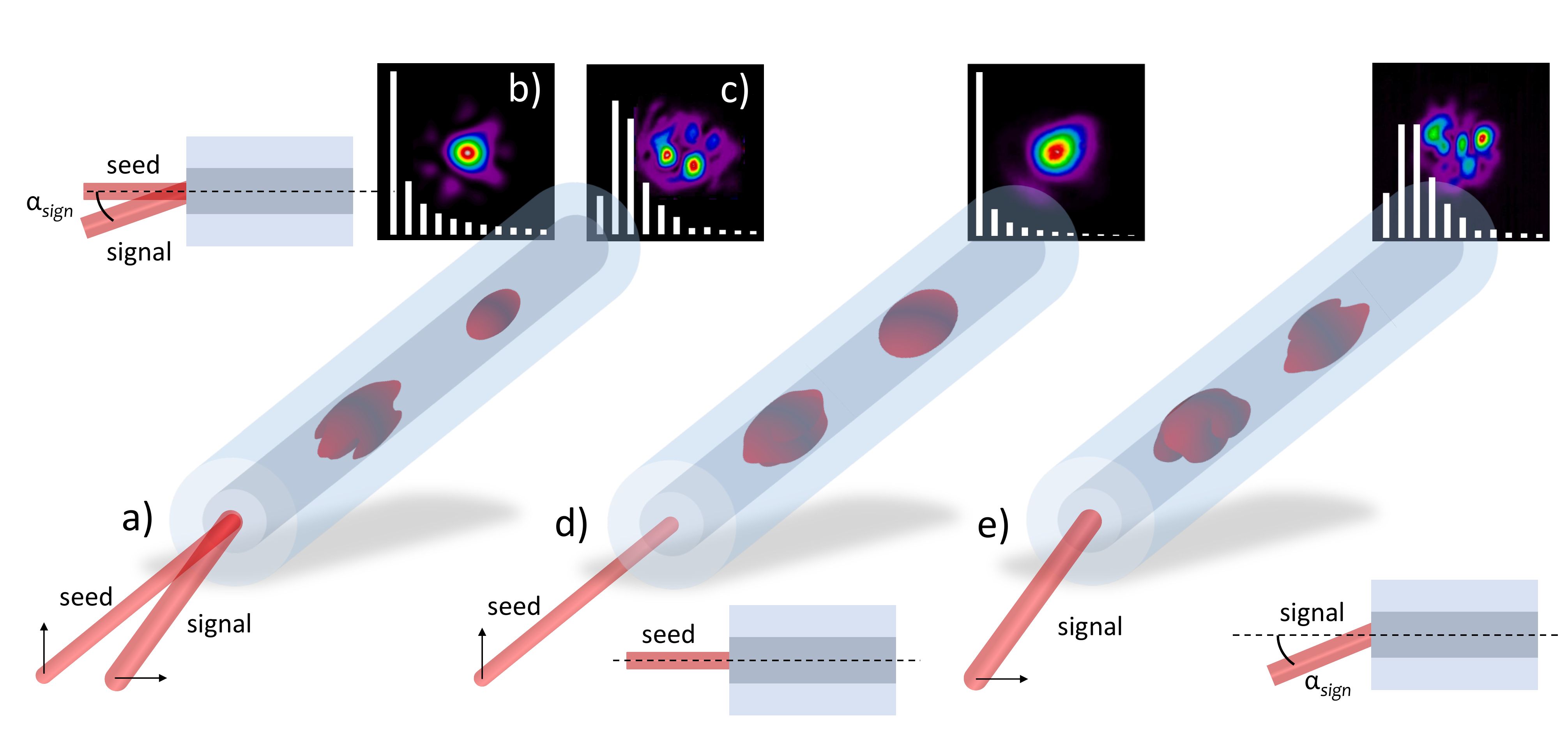}
\caption{(a) Sketch of the MMF-based optical beam-by-beam switching geometry. (b,c) Switch on (b) and off (c) configurations: seed and signal beams are simultaneously injected into the fiber, as shown in (d) and (e), respectively: at the fiber output, the beam emerges with either high (b) or low (c) spatial quality, depending on the seed and signal injection conditions; in each case, the output mode distribution is illustrated by white histograms. (d) Injection of the sole seed beam, straight onto the fiber axis. If its input power is high enough, the output beam self-organizes into a bell shape: its mode distribution is similar to that in (b) (the fundamental mode is most populated); (e) Injection of the sole signal beam with nonzero tilt angle $\alpha_{sign}$. The output beam is speckled, similarly to (c), and the fundamental mode is generally not highly populated. The black arrows represent the polarization vector of the input beams.}
\label{fig:toy}
\end{figure*}

A sketch of the beam-by-beam switching principle is shown in Fig. \ref{fig:toy}a. Consider two orthogonally polarized laser beams, simultaneously injected into the core of a few meter long GRIN MMF. One of the two beams is launched with a tilt angle ($\alpha_{sign} \simeq 2^\circ$) with respect to the fiber axis. This leads to exciting a relatively large number of modes in the fiber. We will refer to such a beam as \emph{signal}. Whereas, a weaker beam, which we dub \emph{seed}, is injected straight onto the fiber axis. Correspondingly, only fewer modes are excited. Now, by progressively increasing the seed power relative to the signal, one observes a cleanup of the total output beam, or a substantial increase of its brightness and quality. When this is achieved, that we will refer to as \emph{switch on} output state, a bell-shaped beam is obtained (see Fig. \ref{fig:toy}b). In this case, the fundamental fiber mode is the most populated in the output beam: the corresponding mode occupancy is sketched by a white histogram in Fig. \ref{fig:toy}b. To the contrary, in the \emph{switch off} configuration, a low spatial quality, speckled beam is observed at the fiber output. Correspondingly, the fundamental mode occupancy is equal or lower than for high-order modes (HOMs). This is shown in Fig. \ref{fig:toy}c. 

In order to better appreciate the difference between the switch-on and switch-off situations, it is useful to consider the limit cases where the power of either the signal or the seed is set to zero. These cases are shown in Figs. \ref{fig:toy}d and \ref{fig:toy}e, respectively. When a single powerful laser beam is injected straight on-axis into the GRIN MMF core, it mostly excites the fundamental mode and a few HOMs. This is a favourable condition for obtaining beam self-cleaning, whereby a bell-like shaped beam is formed at the fiber output (see Fig.\ref{fig:toy}d). Being based on the Kerr effect, beam self-cleaning requires an input peak power which is above a certain threshold value ($P_{thr}$).
Specifically, for a few meters of a standard GRIN MMF, $P_{thr}$ is of the order of a few kWs \cite{krupa2016observation, leventoux20213d}. Now, the value of $P_{thr}$ increases as the injection angle ($\alpha$) grows larger, because a higher number of modes is excited in this case \cite{deliancourt2019kerr}. This significantly limits the possibility of self-cleaning for titled beams: we found that, for $\alpha \simeq 2^\circ$, the value of $P_{thr}$ at $L = 2$ m is so high, that it overcomes of the threshold for fiber damaging. Therefore, in the absence of a seed beam, the titled beam generates a speckled spatial intensity pattern at the fiber output (see Fig. \ref{fig:toy}e), which is similar to that of Fig. \ref{fig:toy}c.

Remarkably, injecting even a weak seed beam along side the highly multimode signal beam permits to all-optically switch the output beam from a low to a high quality state. We found that such optical switching still occurs for seed powers as low as one tenth of the signal power. In other words, one may all-optically (i.e., instantaneously) control the spatial quality of a high-power multimode beam with a relatively weak seed beam. This paves the way for efficient all-optical complex beam processing based on the MMF platform. Fiber lasers stand out among the domains of applications of beam-by-beam switching, given that controlling the spatial quality of their emission still remains a key challenge. Indeed, the switch-on configuration can be seen as a coherent beam combining mechanism that permits the amplification of the seed beam by a multimode signal, which is extremely useful property when building up a laser cavity.

Still, appropriate conditions must be matched for triggering the beam-by-beam switching effect. This will be the main topic of the next Sections. Specifically, we will show that specific constraints must be imposed on the power of both the signal and the seed beam, as well as on the tilting angle $\alpha$. By determining such constraints, we will define the working conditions for the all-optical switch. Importantly, these conditions turn out to be compatible with the operating regime of the state-of-the-art fiber devices. Our results point to the potential of our all-optical switching approach, which could be easily combined within MMF-based technologies, such as space-division-multiplexed communication networks, high-power fiber lasers \cite{tegin2020single}, and high-resolution nonlinear imaging \cite{moussa2021spatiotemporal, wehbi2022continuous}.

This Article is organized as follows: in Section \ref{sec:design}, we numerically provide a proof-of-concept of the all-optical switching effect.
In Section \ref{sec:setup} we describe the experimental setup used for light beam characterization at the MMF output. In Section \ref{sec:exp_limoges} we discuss experimental demonstrations of the switch, including a quantitative assessment of the spatial beam quality. Moreover, we propose a simple and yet extremely useful representation of the experimental results. Specifically, we introduce a geometrical approach-based phase diagram, which allows for discriminating the working regimes (on and off) of the switch. In Section \ref{sec:exp_rome} we directly test the validity of this representation, by carrying out an experimental mode decomposition of the beam at the switch output. Finally, in Section \ref{sec:conclusion} we draw our final conclusions.

\section{Numerical study of the switching effect}
\label{sec:design}

\begin{figure*}[ht!]
\centering\includegraphics[width=16cm]{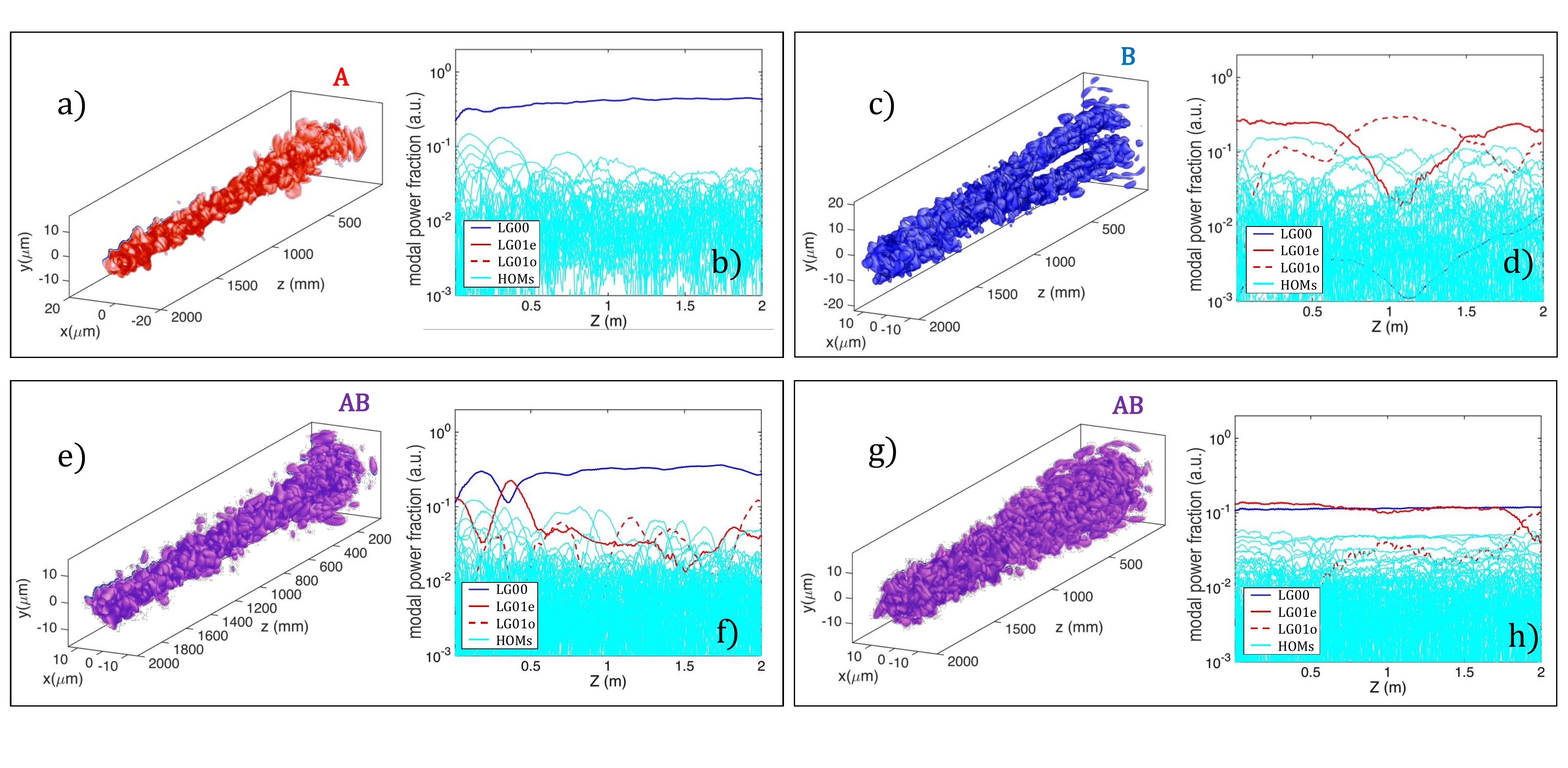}
\caption{Numerical simulations illustrating the beam-by-beam switching principle. a,b) Beam evolution in the presence of the sole input seed beam, with intensity of 5 GW/cm${}^2$ (a); corresponding output mode occupancy (b). c,d) Same as (a,b), with the sole input signal beam, with intensity of 5 GW/cm${}^2$. e,f) Same as (a,b), with both seed and signal beams at the fiber input, each with intensity of 5 GW/cm${}^2$. g,h) Same as (e,f), for weak seed and signal beams (with intensity of 0.5 kW/cm${}^2$, quasi-linear propagation regime.}
\label{fig:simulations}
\end{figure*}

In order to simulate the nonlinear spatial dynamics of a laser beam into a GRIN MMF, we numerically solved the vectorial two-dimensional nonlinear Schr\"odinger equation (2D-NLSE) as in Ref. \cite{krupa2017spatial}
. For sufficiently long optical pulses, and in the normal dispersion regime, nonlinear beam reshaping at the output of a MMF is essentially a spatial effect \cite{krupa2018spatiotemporal}. Therefore, here we do not consider temporal dynamics, i.e., we model our switching principle in the continuous wave (CW) regime. However, as we will see in the following, beam-by-beam switching can also be implemented with short pulsed sources. These are needed in experiments, in order to reach the necessary high peak power values. 

In our simulations, 
we considered a standard parabolic GRIN MMF with a core diameter of 50 $\mu$m, a maximum value for the core refractive index $n_{co}=1.47$, and a cladding refractive index $n_{cl}=1.457$. We took 
into account the presence of fiber disorder, that leads to power exchanges between orthogonally polarized beams.
We considered a non-dispersive Kerr nonlinear index, $n_2=3.2\times10^{-20}$ m$^2$/W$^{-1}$ and the Raman fraction $f_R=0.18$. 
We used an input beam diameter (full-width-half-maximum of intensity) of 40 $\mu$m at a laser wavelength of 1064 nm. 

The results of our numerical simulations are shown in Fig.\ref{fig:simulations}. In particular, in Figs. \ref{fig:simulations}a and \ref{fig:simulations}b, we illustrate the propagation of the sole seed beam (like in Fig. \ref{fig:toy}d). As it can be seen, in the nonlinear regime, the beam progressively shrinks its size: the diameter in the xy plane reduces upon propagation in the z direction (see Fig. \ref{fig:simulations}a). This corresponds to an increase of the fundamental mode population, whose fractional occupation is shown by a dark blue line in Fig. \ref{fig:simulations}b. Whereas the occupancy of HOMs reduces by more than one order of magnitude after 2 m of propagation (light blue curves in Fig. \ref{fig:simulations}b). For the mode representation, we consider the Laguerre-Gauss base, so that the fundamental mode is dubbed LG00, and the first HOM is LG01. The latter is then distinguished into ordinary LG01o and extraordinary LG01e, since in our vector model we took into account fiber birefringence. Specifically, the ordinary index corresponds to the horizontal polarization, i.e., that of the seed, and the extraordinary index corresponds to vertical polarization, i.e., that of the signal.

In Fig. \ref{fig:simulations}b, we present a case that mimics the injection of the mere signal beam. Specifically, we show the nonlinear evolution of a vertically polarized laser beam, which is not injected straight onto the fiber axis. To the contrary, we consider an input beam configuration with a dual-lobe profile (see Fig. \ref{fig:simulations}c). This corresponds to a mode composition at the fiber input where the fundamental mode is poorly populated (see the dark blue line in Fig. \ref{fig:simulations}d). Whereas the most populated mode is the LG01e mode (as shown as a red solid line in Fig. \ref{fig:simulations}d). The relative occupation of LG01e and LG01o modes periodically oscillates during their propagation. Note that the fundamental mode occupancy progressively grows up (dark blue curve in Fig. \ref{fig:simulations}d). However, after 2 m of propagation, its occupancy is still relatively weak, so that the output beam remains speckled, as it occurs in the presence of the sole signal beam. 

Under these conditions, we may reproduce the optical switching effect which occurs whenever both seed and signal beams are simultaneously injected into the fiber core. This is shown in Figs. \ref{fig:simulations}e and \ref{fig:simulations}f, where seed and the signal beams have the same intensity. As it can be seen, the combined beam evolution qualitatively resembles that of Fig. \ref{fig:simulations}a with the seed beam only.
Indeed, the beam progressively shrinks during its propagation (cfr. Fig. \ref{fig:simulations}d), and the fundamental mode is again the most populated at the fiber output (dark blue line in Fig. \ref{fig:simulations}e). As a result, the output beam acquires a bell-like shape, similar to the case reported in the inset of Fig. \ref{fig:toy}b.
Finally, we underline that the optical switching process is actually driven by the MMF nonlinearity. 

As a matter of fact, when considering the seed + signal configuration in the linear propagation regime, the combined beam only slightly reduces its size after two meters of propagation, simply due to the linear guiding properties of the MMF (see Fig. \ref{fig:simulations}g). Moreover, at variance with the nonlinear regime, the occupancy of the fundamental mode remains constant during beam propagation, and there is only a random mode coupling induced exchange of mode occupancy between degenerate modes, as illustrated in Fig. \ref{fig:simulations}h.

\section{Experimental setup}
\label{sec:setup}

Our experiments were carried out by means of two separate experimental setups, involving two different laser sources. Specifically, the source that we used in the first setup was a Nd:Yag laser (Spark Laser, Sirius-type) emitting pulses of 65 ps with 100 kHz repetition rate at 1064 nm. Whereas, in the second setup we used a Yb-based laser (Light Conversion PHAROS-SP-HP), generating pulses with tunable duration, ranging between 174 fs to 7.8 ps with 100 kHz repetition rate, at 1030 nm. 
In addition, our femtosecond laser system was equipped with an optical parametric amplifier, generating an idler beam at 1550 nm, which permitted us to explore the operation of the switch in the anomalous dispersion regime of the fiber as well.
 
\begin{figure}[ht!]
\centering\includegraphics[width=9cm]{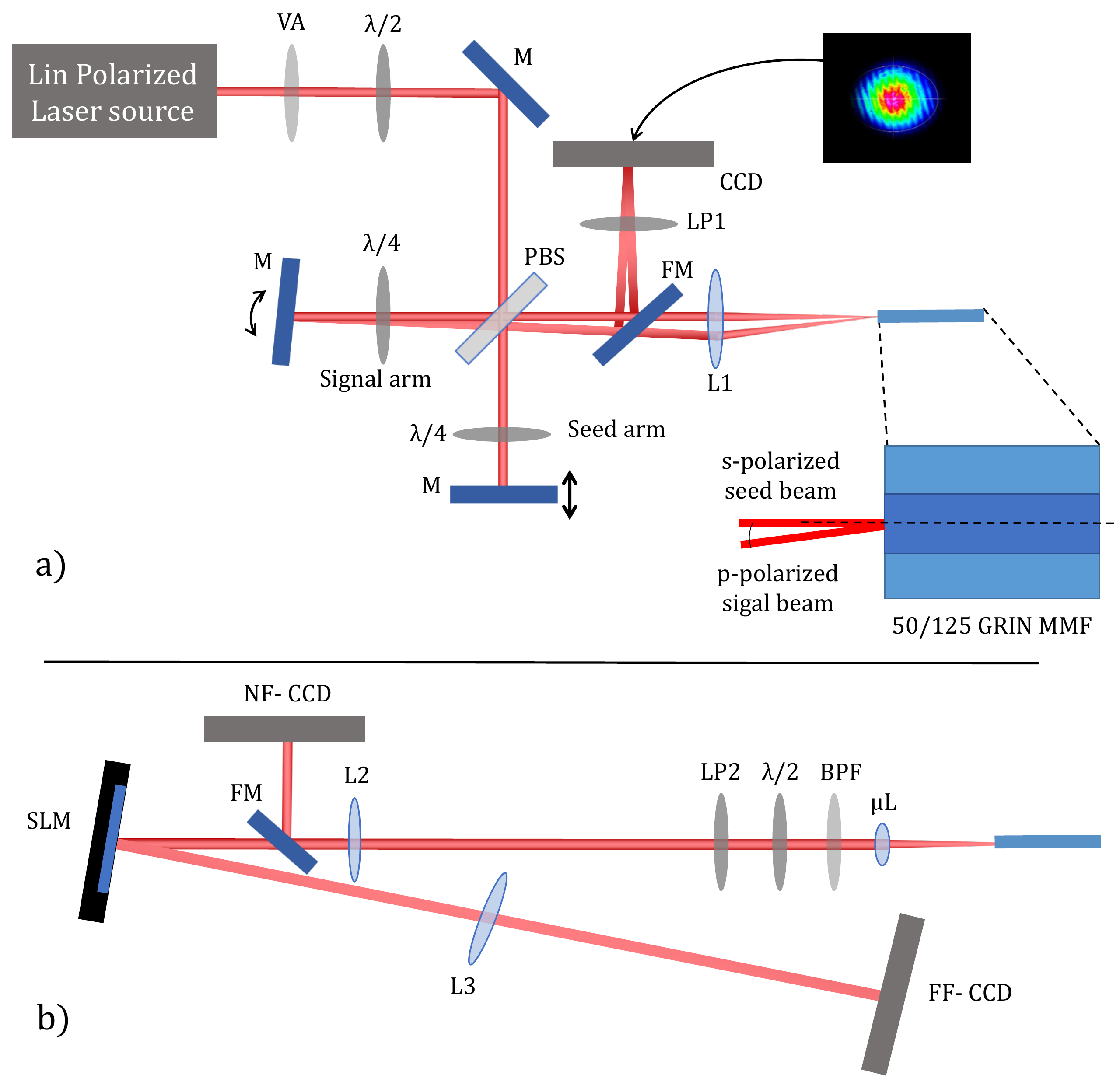}
\caption{Sketch of the experimental setup; (a) excitation setup; (b) spatial analysis setup. VA: Variable Attenuator; $\lambda/2$: Half waveplate; $\lambda/4$: Quarter waveplate; M: Mirror; FM: Flipping Mirror; PBS: Polarizing Beam Splitter; FM: Flipping Mirror; LP: Linear Polarizer; L1: Injection Lens; $\mu$L: Microlens; BPF: Band Pass Filter; L2, L3: Mode decomposition lenses; SLM: Spatial Light Modulator.  
}
\label{fig:set-up}
\end{figure}

\begin{figure*}[hb!]
\centering\includegraphics[width=16cm]{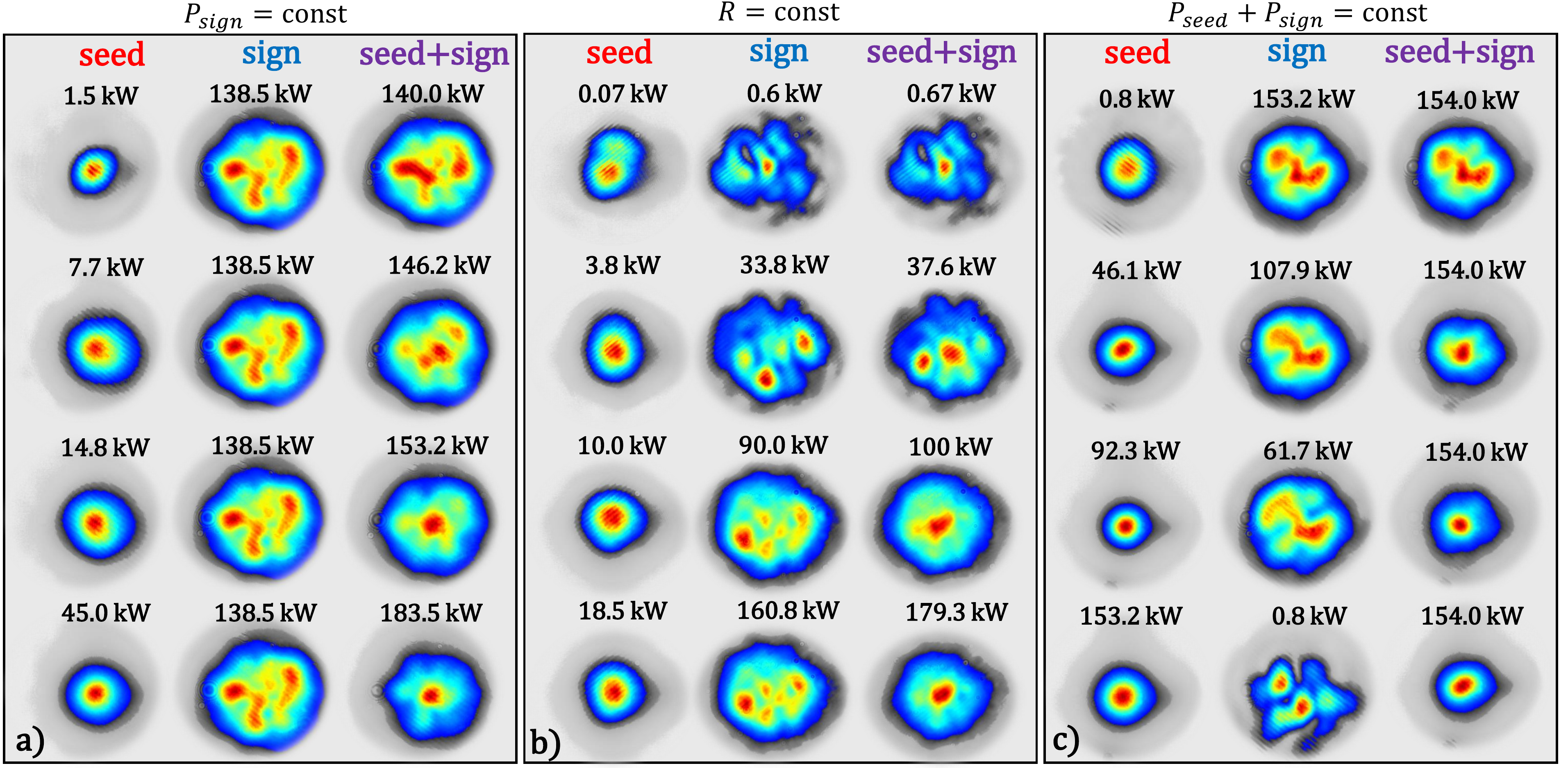}
\caption{Beam images at the fiber output for three separate beam switching experiments. Here we fix either: the power of the signal beam (a), the ratio between the power of the seed and the signal beam $R\simeq 1:9$ (b), and the sum of the latter (c). In each panel, the first (second) column shows the case when the sole seed (signal) is injected; for the third column, both beams are injected.}
\label{fig:exp_lim}
\end{figure*}

The two setups were rather similar, at least, as far as the optical components placed before the fiber input (as sketched in Fig. \ref{fig:set-up}a) are concerned. We split the laser source into two temporally overlapping laser beams with orthogonal linear polarization, in order to inject them into a 2 m long GRIN MMF with 50 $\mu$m core diameter, according to the geometry of Fig. \ref{fig:toy}a, and with $\alpha_{sign} = 1.83^\circ$. The splitting of the laser source beam into two arms (seed and signal), as well as the control of their state of polarization, was obtained by means of a polarizing beam splitter (PBS). Specifically, we used a configuration that resembles that of a Michelson interferometer. However, at variance with the latter, the presence of a PBS, on the one side, produces two beams with orthogonal linear polarization and, on the other side, requires the insertion of quarter-waveplates ($\lambda$/4) in both the seed and the signal arms. In fact, in the absence of such $\lambda/4$ retarders, both seed and signal beams would be reflected back towards the laser source by the PBS, since they would maintain their polarization state. To the contrary, the combined action of the PBS, the $\lambda$/4 and the mirrors (M) allows for light to reach the fiber input. In particular, the mirror of the signal arm was slightly tilted, in order to introduce a nonzero injection tilt angle $\alpha_{sign}$ as in Fig. \ref{fig:toy}. Note that, by acting on the orientation angle of each $\lambda/4$, we could vary the power of each arm, without varying that of the other (see results in Fig. \ref{fig:exp_lim}a). Whereas, by acting on the half-waveplate ($\lambda$/2), we could vary the value of the power fraction $R=P_{seed}/P_{sign}$, while keeping constant that of $P_{seed} + P_{sign}$ (as it occurs for the results shown in Fig. \ref{fig:exp_lim}b and in Fig. \ref{fig:mode_decomp}, respectively). The latter, instead, was varied by means of a variable attenuator (VA), as in Fig. \ref{fig:exp_lim}c. 

Finally, as we were working with short pulses, in order to verify the synchronization of the seed and signal beams, we interposed a flipping mirror (FM) between the PBS and the injection lens (L1) which focuses both beams into the fiber core. A CCD camera was placed after an optical path, which was equivalent to that travelled by the light in order to reach the fiber input. We inserted a $45^\circ$ oriented polarizer (LP1) in order to image interference fringes on the camera. In this way, the temporal synchronization of the two beams could be verified by the occurrence of an interference pattern. To this purpose, the mirror of the seed beam was placed on a micrometric slit, as it is conventionally done for standard Michelson interferometers.

The main difference between the two setups consists in the methods for beam analysis at the fiber output. In the setup involving the Nd:Yag laser, we only measured the intensity profile of the output beam by means of a silicon CCD camera (Gentec Beamage), along with the fiber power transmission. Whereas, the setup that involves the Yb-based laser system also included an apparatus for performing the mode decomposition (MD) of the output beam. A sketch of the optical components that we used for the MD study is illustrated in Fig. \ref{fig:set-up}b. 

At the fiber output, the beam is collected by means of a microlens ($\mu L$). A second lens (L2) is used for imaging the near-field at the fiber output onto a spatial light modulator (SLM) (Hamamatsu LCOS- X15213). In order to image the field that reaches to the SLM, we inserted a flipping mirror (FM2), so that the focus of the L2 lens coincides with the screen of a CCD camera (Gentec Beamage-4M-IR), dubbed as near-field (NF) camera. Finally, a third lens (L3) projects the Fourier transform of the beam reflected at the SLM surface onto a CCD camera (Gentec Beamage-4M-IR), dubbed as far-field (FF) camera. Note that, since the SLM only works with a linearly polarized beam, we placed a linear polarizer (LP2) right before L2. In order to maximize the power that reaches the SLM, we also added a $\lambda/2$ waveplate right before LP2. We underline that the orientation angle of the $\lambda/2$ waveplate was different for the three configurations shown in Fig. \ref{fig:toy}, since the output polarization depends on the input state of polarization. Moreover, a bandpass filter (BPF) was used in order to avoid a possible loss of temporal coherence due to significant self-phase-modulation induced spectral broadening in the MMF. This would be disturbing for the MD analysis, since the SLM requires to operate under nearly monochromatic excitation.

The MD analysis relies on the association to each intensity pattern on the NF camera to a set of images on the FF camera, which are obtained by varying the phase mask that is encoded by the SLM. Such phase masks are chosen in order to separate information coming from each of the modes that compose the output beam. Thus, by means of a numerical algorithm, it is possible to retrieve the value of amplitude and phase associated to each mode for a given output beam. In this way, we could reconstruct the beam patterns shown in Figs. \ref{fig:mode_decomp}b-d. Here, we do not provide details about our MD algorithm, whose full description can be found in Ref. \cite{gervaziev2020mode}. 

As a final note, it is important to underline that the presence of the linear polarizer LP2 allows for properly measuring the mode content of a beam which has a linear polarization state. This idea applies well to the cases of either sole seed or sole signal injection, where the output state of polarization remains approximately linear even at the MMF output. Specifically, besides the occurrence of some depolarization effects, the circular component of the output beam polarization was negligible with respect to the linear one. To the contrary, one may wonder what is the polarization state at the fiber output, in the case of simultaneous injection of both seed and signal. However, we verified that, in the experimental conditions of Fig. \ref{fig:mode_decomp}, the degree of linear polarization at the fiber output still remained quite higher than that of circular polarization. Moreover, the direction of the linear polarization vector, i.e., the angle of the $\lambda/2$ that maximized the power transmitted through $LP2$, was the same as that corresponding to the injection of the sole seed beam. Therefore, for all of the experiments reported in this work, the MD performed along one polarization direction can be identified, within a reasonable approximation, with that corresponding to the entire output beam.

\section{Experimental results with the Nd:Yag laser}
\label{sec:exp_limoges}

Fig. \ref{fig:exp_lim} illustrates the experimental demonstration of MMF-based beam-by-beam cleanup. Here we report three different experiments, each with a different set of initial conditions. Specifically, we fixed: (a) the power of the signal beam (cfr., Fig. \ref{fig:exp_lim}a); (b) the ratio $R$ between the seed and the signal power (cfr., Fig. \ref{fig:exp_lim}b); (c) or the sum of the seed and the signal power (cfr. Fig. \ref{fig:exp_lim}c). In each subfigure, we show the beam profile in the presence of: (i) the sole seed beam (left column); (ii) the sole signal beam (mid column); (iii) both seed and signal beams (right column), as the power of the seed grows larger. Note that we associated to each beam profile the value of the peak power at the fiber input, which was calculated as
\begin{equation}
    P = \frac{\langle P \rangle}{\tau \cdot RR},
    \label{eq:pin}
\end{equation}
where $\langle P \rangle$ is the measured average beam power at the fiber output, RR is the laser repetition rate, and $\tau$ is the pulse duration.

As it can be seen in Fig. \ref{fig:exp_lim}a, whenever the signal beam power is fixed, the seed beam must be powerful enough in order to cleanup the overall output beam. Note that here the seed power threshold value ($P_{seed}$ = 14.7 kW) is about ten times higher than the power threshold for self-cleaning of the seed beam alone, i.e., $P_{seed}$ = 1.54 kW. Moreover, the same panel also shows that signal cleanup (switch-on condition) occurs for a power ratio ($R$) between seed and signal as low as $1:9$ (i.e., 14.7 kW : 138 kW). 
As a matter of fact, in Fig. \ref{fig:exp_lim}b we show that when R is fixed (e.g., $R=1:9$), there is again a threshold value of $P_{seed}$ for obtaining cleanup of the signal beam: here this threshold turns out to be equal to 10 kW. The results in Figs. \ref{fig:exp_lim}a and \ref{fig:exp_lim}b demonstrate that one must control both $P_{seed}$ and $R$, in order to properly operate the beam-by-beam switching. 
Finally, in Fig. \ref{fig:exp_lim}c, we show that results obtained when the sum of the seed and signal power is kept at a constant value are consistent with the observations in Figs. \ref{fig:exp_lim}a,b. 

It is important to underline that, since seed and signal have orthogonal states of polarization, when both beams are simultaneously coupled into the fiber core, the total peak power in the fiber is
\begin{equation}
    P_{eff} = P_{seed} + P_{sign}.
\label{eq_P_sum}
\end{equation}
This is coherent with the peak power values in Fig. \ref{fig:exp_lim} since, in all panels, the power value in the third column is equal to the sum of the powers appearing in the first two columns.


\subsection*{Quantitative evaluation of the spatial quality}

When discussing the experimental results in Fig. \ref{fig:exp_lim}, we only qualitatively distinguished the shape and quality of the output beam profiles. Namely, we did not yet quantify the spatial quality of the output beams. Here, we provide a quantitative analysis of the images reported in Fig. \ref{fig:exp_lim}. Specifically, we evaluate the correlation of the intensity of the output beam ($I$) with that of the fundamental mode of the GRIN MMF that we used in the experiments ($I_0$). According to the specifications provided by the manufacturer, the fundamental mode intensity profile can be represented in terms of a 2D Gaussian beam shape with a waist $w$ = 6.33 $\mu$m. Hence, the image correlation with the fundamental mode is evaluated as \cite{deliancourt2019kerr}

\begin{equation}
    Cor = \frac{\int I(x,y) I_0(x,y) dx dy}{\sqrt{\int I(x,y)^2 dx dy\int I_0(x,y)^2 dx dy}},
    \label{eq:cor}
\end{equation}

where $I_0 \propto \exp\{-2{(x^2+y^2)}/{w^2}\}$. Note that both $I$ and $I_0$ are normalized in a way that their integral over the CCD camera area is equal to 1. The values of $Cor$ which correspond to the same experimental conditions as in Figs. \ref{fig:exp_lim}a-c are reported in Figs. \ref{fig:couples}a-c. For sake of completeness, we calculate $Cor$ for all of the three configurations that we have considered in Fig. \ref{fig:exp_lim}, i.e., injection of the sole seed, of the sole signal, and of both seed and signal. The corresponding results are collected in Fig. \ref{fig:couples}a-c, and marked by red, blue and violet dots, respectively. Therein, we plot as void circles data that correspond to low spatial quality output beams in Fig. \ref{fig:exp_lim}. 

Whereas, beams that we dubbed as highly spatial quality or clean beams are represented here by full circles. As it can be seen in all of the graphs of Fig. \ref{fig:couples}a-c, there is a threshold value of the correlation ($Cor$ = 0.68), which permits to split apart void and full circles. As a consequence, we can associate to an output beam the label of either "high" or "low" spatial quality, by computing its correlation $Cor$ with the fundamental from Eq. (\ref{eq:cor}). If $Cor >(<)$ 0.68, then we consider the corresponding beam to be of high or low quality, respectively.

As a final note, we stress the reason behind the evaluation of the spatial beam quality by means of $Cor$ instead of the more conventional $M^2$ parameter. Since in the next Section, we will compare the evaluation of the beam quality with MD experiments, the evaluation of $Cor$ is more appropriate than that of $M^2$. In fact, the main difference between $Cor$ and $M^2$ is that the former is somewhat related to the occupancy of the fundamental mode, i.e., it is bound to the fiber size, whereas the latter does not depend on the beam diameter (at least as long as collimated beams are considered) as it accounts for the divergence of laser beams that propagate in the free-space. 

\subsection*{Geometrical approach-based representation}

\begin{figure*}[ht!]
\centering\includegraphics[width=16.5cm]{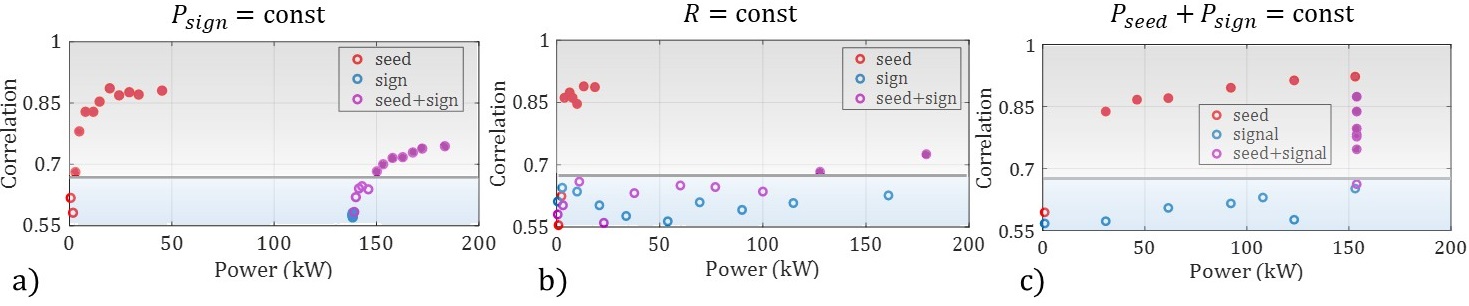}
\centering\includegraphics[width=16.5cm]{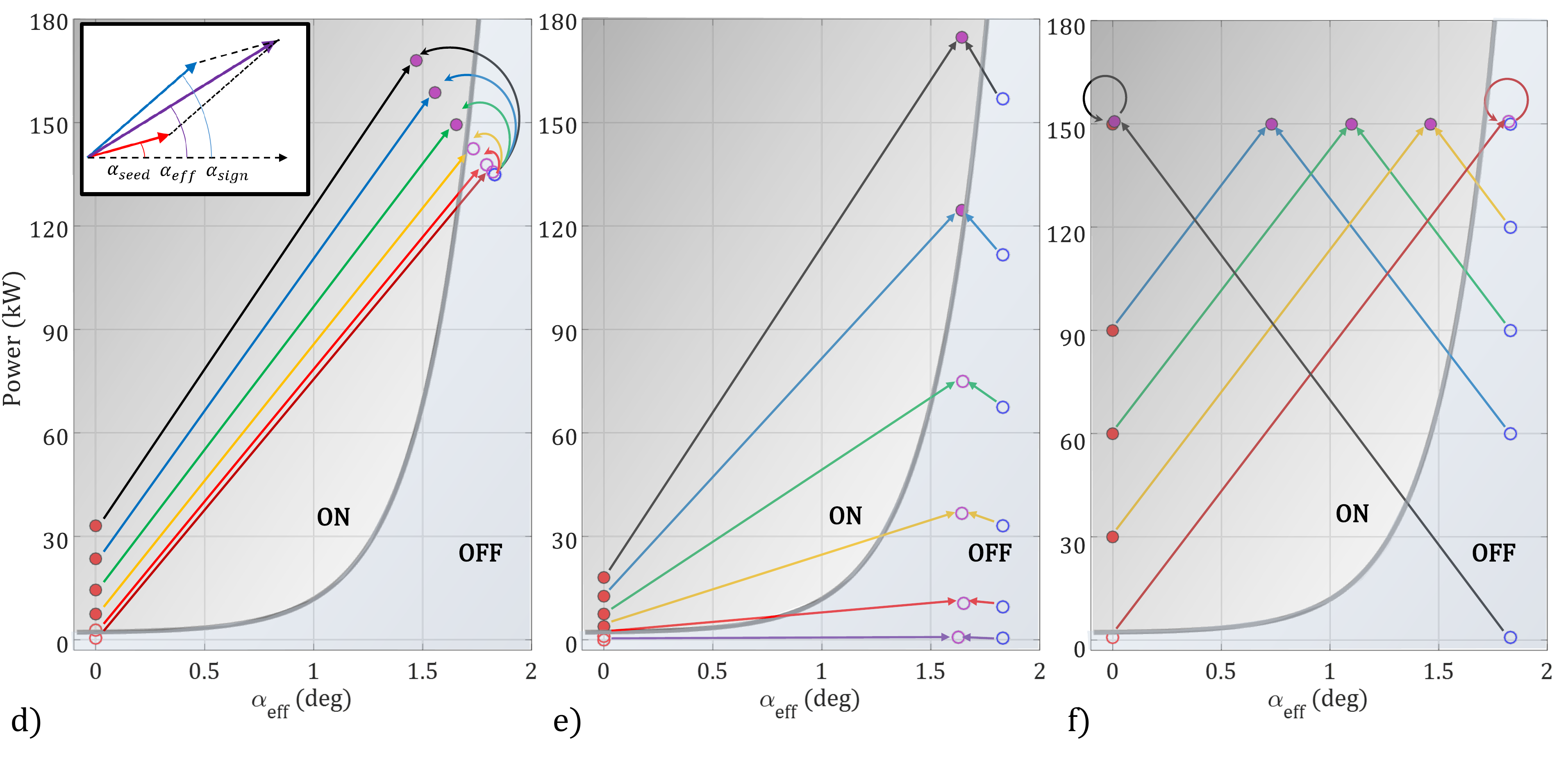}
\caption{a-c) Evaluation of the correlation parameter $Cor$, calculated by Eq. (\ref{eq:cor}), for the experimental data in Figs. \ref{fig:exp_lim}a-c. d-f) Geometrical-based representation of the data in Figs. \ref{fig:exp_lim}a-c, as depicted in the inset of panel (d). The values of power in the vertical axis are calculated by Eq. (\ref{eq_P_sum}), whereas the values of $\alpha$ in the horizontal axis are calculated by Eq. (\ref{eq:alpha}).}
\label{fig:couples}
\end{figure*}
Here, we introduce a representation of the experimental data presented in Fig. \ref{fig:couples}a-c, which relies on a purely geometrical approach. Specifically, we describe the configuration where both seed and signal beams are injected together into the fiber in terms of the sum of two vectors. Each vector has a modulus equal to the beam power, whereas the orientation of the vector is measured by its injection angle ($\alpha$) with respect to the fiber axis. We sketch the geometry of such representation in the inset of Fig. \ref{fig:couples}d. 
Since we are dealing with small angles (smaller than 2${}^\circ$), we can consider the approximations $\cos{\alpha} \simeq 1$, $\sin{\alpha} \simeq\tan{\alpha} \simeq \alpha$. Thus, by following the sketch in the inset of Fig. \ref{fig:couples}d, we may write
\begin{equation}
    \alpha_{eff} = \frac{P_{seed}\cdot \alpha_{seed}+P_{sign}\cdot \alpha_{sign}}{P_{seed} + P_{sign}}. 
    \label{eq:alpha}
\end{equation}
We underline that the sum in Eq.(\ref{eq:alpha}) does not correspond to a vectorial sum of the electric fields associated with the seed and the signal beams, which are orthogonally polarized. Indeed, here we are performing a sum of two vectors, whose modulus is given by the beam power instead of the field amplitude. In particular, in our case, we have $\alpha_{seed} = 0$.

The validity of such a geometric representation is strengthened by the following argument. As discussed in the introductory Section, it is well-known that, at a given propagation distance, the threshold power for self-enhancing the spatial quality of laser beam propagating into a MMF (or beam self-cleaning) depends on $\alpha$. Specifically, such a threshold power increases as $\alpha$ grows larger, as reported in Ref. \cite{deliancourt2019kerr}. Therefore, it is reasonable to associate an effective angle to the seed + signal input configuration, which is obtained as the weighted average of the injection angles of seed and signal beams. Here the weights are proportional to the power of each beam. As a matter of fact, this is in agreement with Eq.(\ref{eq:alpha}), where the weights for obtaining the average or effective angle are exactly provided by the power of each beam. 

Within this representation, we may display the experimental data of Fig. \ref{fig:exp_lim} in terms of ($\alpha$,$P$) phase plane diagrams. The result is shown in Figs. \ref{fig:couples}d-f. As it can be seen, for each couple of points (seed, signal) which is depicted by either red or blue circles, respectively, we may associate a violet dot, whose coordinates ($\alpha$ and $P$) are calculated by means of Eqs. (\ref{eq:alpha}) and (\ref{eq_P_sum}), respectively. Each association is marked by two colored arrows. Analogously to Fig. \ref{fig:couples}a-c, again we assigned full and empty circles to high or low spatial quality beams, respectively. 

As it can be seen, all the graphs in Figs. \ref{fig:couples}d-f can be split in two separate phase plane domains. A first zone, in the bottom-right side of all graphs, collects all of the void circles, and it can be associated with the switch-off regime of Fig. \ref{fig:toy}c. Whereas, a second region, in the top-left side of the panels in Figs. \ref{fig:couples}d-f, corresponds to the switch-on condition of Fig. \ref{fig:toy}b, since it incorporates all of the full circles.

The switch-on and switch-off regions are separated by a threshold condition, that links the input peak power to the injection angle. Such a threshold is depicted by a gray solid line in Figs. \ref{fig:couples}d-f. To the best of our knowledge, no studies of the shape of such curve have been reported in the literature so far. Here we used an empirical exponential curve, which appears to properly describe the separation between the switch-on and switch-off conditions, as we will also see in the following Section.

It is worth to underline that, although the quantitative evaluation of the beam quality turns out to be effective for representing the cross-cleaning process, upon which the optical switching effect is based, the exact value of $Cor$ which provides the partition of the ($\alpha, P$) plane is somehow arbitrary. As a matter of fact, the threshold value $Cor = 0.68$ is chosen by starting from a qualitative distinction between low and high quality beams in Fig. \ref{fig:exp_lim}. Hence, the gray separatrix in Fig. \ref{fig:couples}d-f does not describe for a sort of phase-transition from low to high-quality beams. Indeed, cross-cleaning relies on the same physical mechanisms (Kerr effect) as beam self-cleaning does. The latter only occurs when the input coupling conditions are favorable enough (in terms of peak power and injection angle, respectively). However, the occurrence of beam self-cleaning does not have a precise power threshold, so that suddenly it appears when a certain input power and coupling condition is reached. To the contrary, it is a phenomenon that leads a beam towards its state of equilibrium (bell-shape), which is gradually reached upon its propagation \cite{mangini2022statistical}.

\section{Experimental results with the Yb-laser}
\label{sec:exp_rome}

In this Section, we report experiments carried out with the Yb-laser source at 1030 nm; we have set the pulse duration to 2 ps, and varied the values of the signal injection angle $\alpha_{sign}$. With this setup, we have also carried out an experimental MD analysis of the output beam, for all the beam configurations that we have described in the previous Section, i.e., with the injection either the sole seed, of the sole signal, or both. Finally, to further extend the regime of operation of the beam-by-beam switching principle, we also compare experiments carried out with femtosecond pulses in either the normal (1030 nm) or in the anomalous (1550 nm) dispersion regime, respectively.


\begin{figure}[ht!]
\centering\includegraphics[width=9cm]{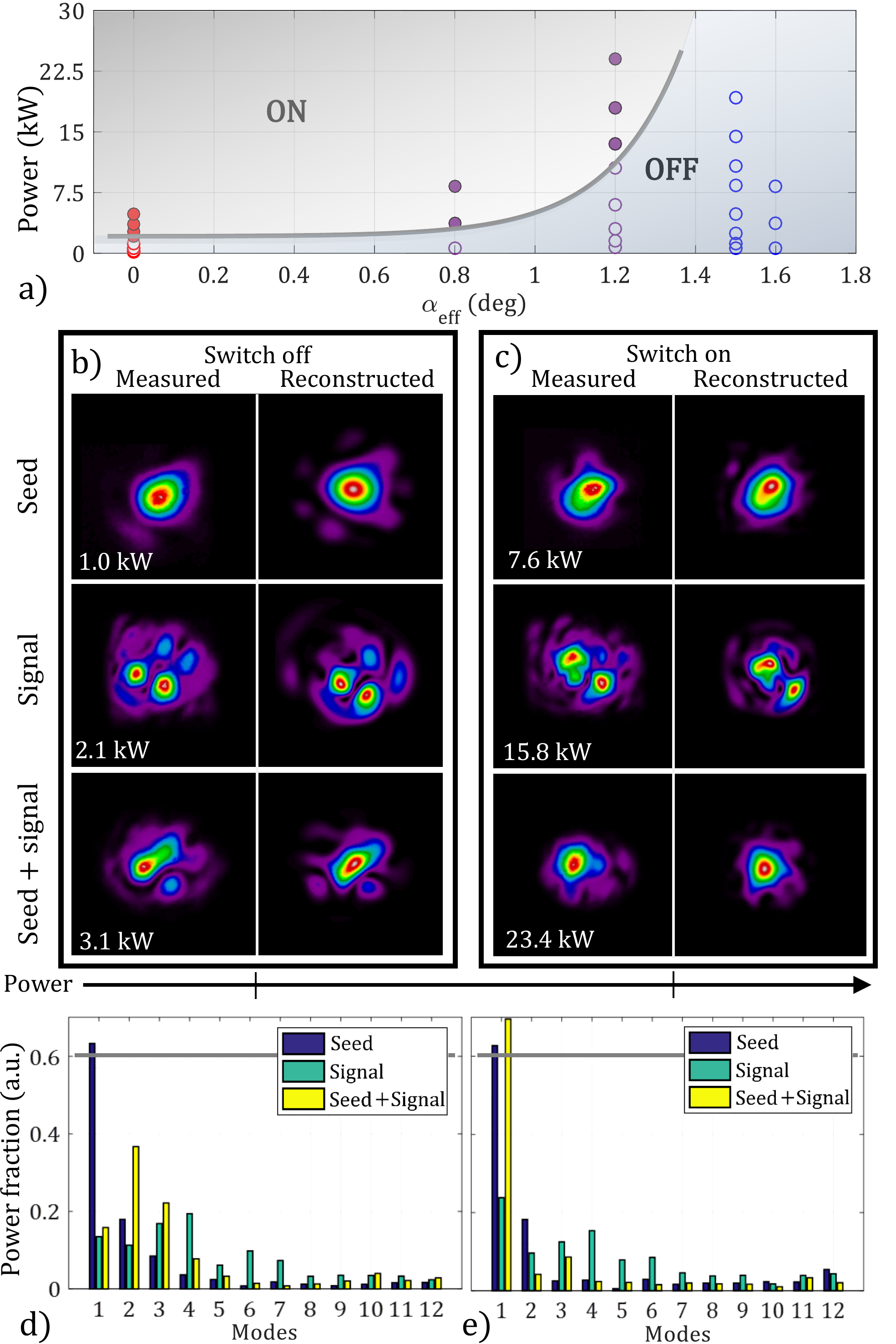}
\caption{Experimental results with the Yb-laser. a) Phase plane diagram of the switch operation, as in Fig. \ref{fig:couples}d-f. Two set of data, corresponding to $\alpha_{seed} = 1.5^\circ$ or $\alpha_{seed} = 1.6^\circ$, respectively, are plotted on the same graph, in order to highlight the general validity of the separatrix (gray solid line) between the switch-on and the switch-off regions. b,c) Measured output beam profile (left column) and its reconstruction by means of the MD analysis (right column). In these experiments, the ratio between the seed and the signal power is equal to R = 1:2; the power of the seed beam was varied from 1.0 kW (b) up to 7.6 kW (c). 
d,e) MD results corresponding to the output beams in (b,c).}
\label{fig:mode_decomp}
\end{figure}

\subsection*{Experimental results with different values of $\alpha_{sign}$}
At first, we repeated our experiments with input beam conditions similar to the experiments with the Nd:Yag laser. Here we used a length $L$ = 3 m of the same GRIN MMF; we set $R$ = 1:2, and considered two different values of $\alpha_{sign}$, i.e., $\alpha_{sign} = 1.5^\circ$ and $\alpha_{sign} = 1.6^\circ$, respectively. The overall results are summarized in the phase plane portrait of Fig. \ref{fig:mode_decomp}a, as we did in Figs. \ref{fig:couples}d-f. As it can be seen, even when varying the injection angle of the signal, we always obtain an exponential dependence of the threshold power on angle $\alpha_{sign}$, which separates the switch-on and the switch-off regions.

One may notice that the exponential separatrices in Fig.\ref{fig:mode_decomp}a and in Figs. \ref{fig:couples}d-f do not exactly overlap. This might be due to differences in the experimental conditions, i.e., the different bending of the MMF, as well as its length (in the Nd:Yag laser experiments we used $L$ = 2 m, and here $L$ = 3 m). However, one may appreciate that, in the spite of the different wavelengths and pulse durations of the two laser sources, the threshold input peak powers remain relatively similar in both two experiments. This is consistent with previous observations that beam self-cleaning is a purely spatial effect while group velocity dispersion remains negligible \cite{mangini2022statistical}.

\subsection*{Mode decomposition}

In order to complement the measurements of the correlation between the output intensity and that of the fundamental mode with a more comprehensive information about the modal content of the output field, we carried out the MD of beam at the fiber output. We considered the case with $\alpha_{sign} = 1.6^\circ$ (cfr. Fig. \ref{fig:mode_decomp}a). In our MD, we computed the power fraction in the fundamental mode, i.e., the LG00 mode, along with that of the first 77 HOMs. Such relatively large number of modes was found to provide a good compromise between the time convergence of the MD algorithm, and the quality of the near-field reconstruction of the output intensity profiles. Indeed, in Figs. \ref{fig:mode_decomp}b,c, one may appreciate the similarity between the measured output beam profile (left column) and its reconstruction by means of the MD technique (right column). Specifically, in the subpanels, we compare the images obtained with: (i) injection of the sole seed (top row); (ii) injection of the sole signal (mid row); (iii) injection of both signal and seed (bottom row). In addition, in Figs. \ref{fig:mode_decomp}b,c, we compare the results for two values of $P_{seed}$, i.e., 1.0 kW and 7.6 kW, which correspond to the switch-off and switch-on cases, respectively. 

The MD results for the cases in Figs. \ref{fig:mode_decomp}b,c are shown in Fig. \ref{fig:mode_decomp}d,e. In order to improve the readability of the figures, here we report the measured relative mode occupancy of the first 12 modes only. In this regard, it is worth underlying that the mode occupancy quickly decreases as the mode number grows larger. 
The most evident feature of Fig. \ref{fig:mode_decomp}d,e is that, whenever the output beam has a bell shape, i.e., for the seed alone and for the total field with $P_{eff}$ = 23.4 kW, 
the fundamental is the most populated mode. Conversely, whenever the fundamental mode has an occupancy close to that of some HOMs, a speckled output beam intensity profile is obtained.

From the MD analysis, we may empirically define a threshold value of the relative fundamental mode occupancy, which permits for distinguishing between high and low quality beams. Such a value turns out to be equal to 0.6, and it is indicated by an horizontal line in Figs. \ref{fig:mode_decomp}d,e. This result further supports the validity of the geometrical representation that we have introduced in Section \ref{sec:exp_limoges}. In fact, it turns out that evaluating the correlation of a beam with the fundamental mode by means of Eq. (\ref{eq:cor}) also provides a good tool for distinguishing among high and low quality beams. This is because crossing the threshold for $Cor$ also corresponds to crossing the fundamental mode occupancy threshold.

\subsection*{Switch operation in the femtosecond pulse regime}

\begin{figure}[ht!]
\centering\includegraphics[width=8.9cm]{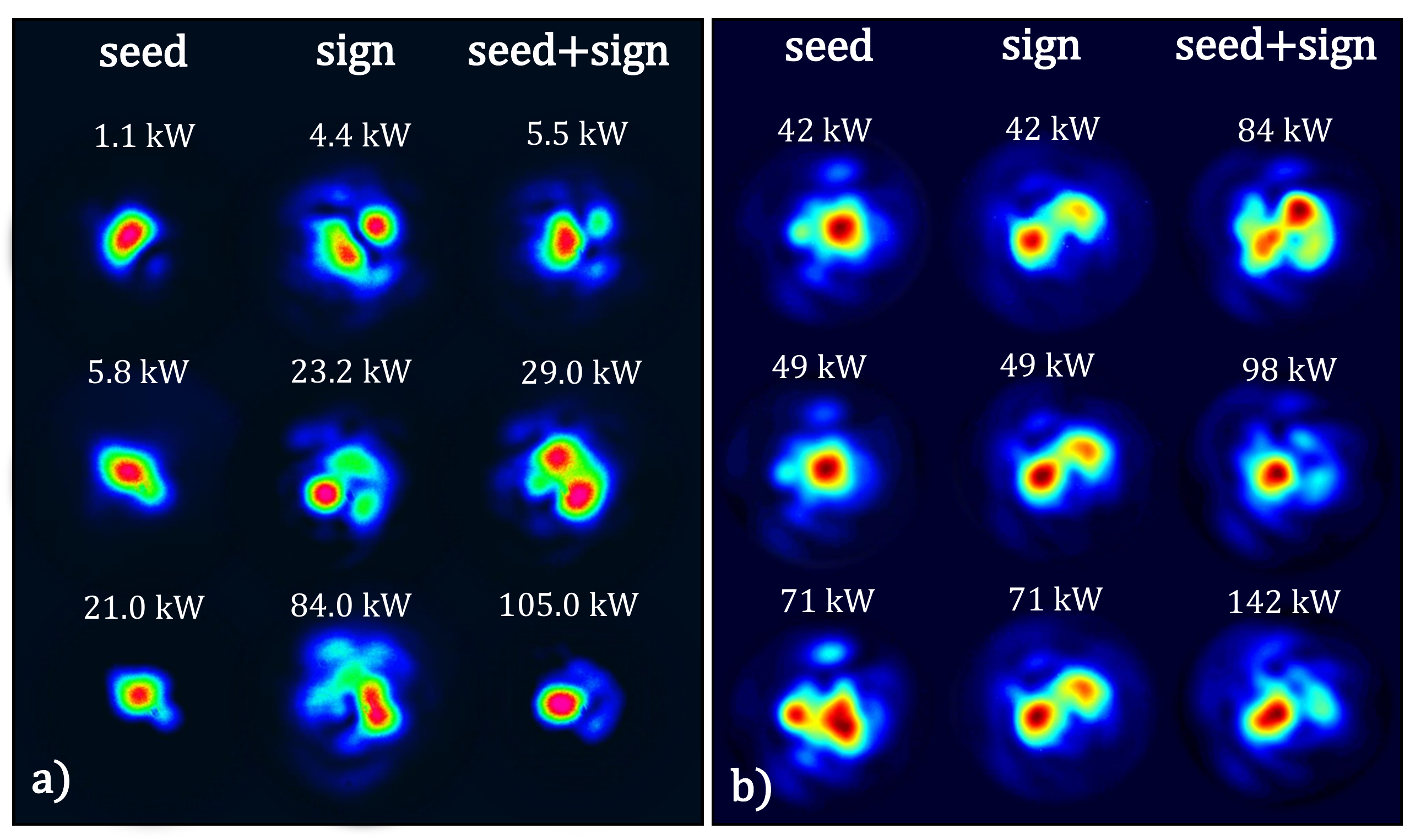}
\caption{Ultrafast beam-by-beam control using orthogonally polarized femtosecond pulses. a) 174 fs pulses at 1030 nm (normal dispersion regime), input power ratio 1:4. b) 70 fs pulses at 1550 nm (anomalous dispersion regime), input power ratio 1:1.}
\label{fig:exp_femto}
\end{figure}

In this final section, we experimentally tested the possibility of extending the operation of the beam-by-beam cleanup principle even in the ultrashort pulse regime. Specifically, we used femtosecond pulses propagating in either the normal or in the anomalous dispersion regime: the corresponding results are shown in Figs. \ref{fig:exp_femto}a and \ref{fig:exp_femto}b, respectively. We underline that, whenever intense femtosecond pulses are used, significant nonlinear spectral broadening due to self-phase modulation (SPM) may take place. Kerr-induced beam cleanup involves beam reshaping which occurs in a limited bandwidth around the input wavelength. Therefore, when injecting pulses either in the normal or in the anomalous dispersion regime, at the fiber output we placed a bandpass filter centered at either 1030 nm or 1550 nm, respectively, with a 5 nm bandwidth. In particular, in the anomalous dispersion regime SPM leads to soliton formation at input peak power values slightly higher that the threshold for beam cleanup. Since femtosecond solitons are subject to Raman-induced frequency red-shift, their wavelength rapidly escapes from that of the pump (1550 nm) \cite{zitelli2021single}. This limits the possibility of obtaining beam cleanup in the anomalous dispersion regime to a narrow range of input powers, i.e., for input power values between the cleanup and the soliton thresholds, respectively. This can be seen in Fig. \ref{fig:exp_femto}b: the seed beam loses its bell-shape when the input peak power is increased from 49 kW to 71 kW. 

Nevertheless, by comparing Fig. \ref{fig:exp_femto}a (normal dispersion regime) with Fig. \ref{fig:exp_femto}b (anomalous dispersion regime), one can see that the all-optical beam switching principle also works in the femtosecond pulse regime. Moreover, the results in Fig. \ref{fig:exp_femto} show that the beam-by-beam cleanup effect can be obtained, regardless of the sign of chromatic dispersion. Interestingly, Fig. \ref{fig:exp_femto}b indicates that all-optical multimode beam control can also be obtained with ultrafast pulses in the telecom window, which could be exploited for ultrafast signal gating and demultiplexing functionalities.

\section{Conclusion}
\label{sec:conclusion}

In conclusion, we introduced and experimentally demonstrated the all-optical control of multimode beams in optical fibers. Specifically, a relatively weak seed beam may control the spatial quality of an intense, orthogonally polarized signal beam at the fiber output. The seed beam can lead to a substantial enhancement of the spatial quality of signal beams with up to tenfold higher power, thanks to a proper adjustment of their injection conditions at the fiber input. The operating mechanism of the switch relies on the nonlinear properties of GRIN MMFs, and, in particular, on the Kerr beam cleaning effect in GRIN fibers. 
Numerical simulations, based on the vector 2D-NLSE, predict that the propagation of two orthogonally polarized laser beams with different multimode content in a GRIN MMF may lead to the enhancement of the overall beam quality at the fiber output, leading to the generation of a bell-shaped profile with a waist close to that of the fundamental mode of the fiber. This concept was experimentally verified, and the beam injection configurations leading to the beam-by-beam cleanup effect, i.e., the switch-on and switch-off conditions, could be determined. A simple empirical geometrical representation was proposed, and experimentally validated across an extensive range of different input beam conditions, laser wavelengths and pulse durations. The output beam quality was quantitatively assessed by computing the intensity correlation of the output images with the intensity profile of the fundamental mode. Finally, we confirmed our findings by carrying out a mode decomposition analysis of beams emerging from the fiber output. This confirmed that, indeed, cleaned beams have a dominant fundamental mode contribution.

Because of its all-in-fiber nature, beam-by-beam cleaning can be naturally incorporated in state-of-the-art MMF-based technologies, such as space division multiplexed networks and high-power fiber lasers. Finally, since the beam self-cleaning effect has been described as result of wave thermalization \cite{wu2019thermodynamic, mangini2022statistical}, our results may pave the way for new developments in the physics of multimode wave systems.

\bibliographystyle{IEEEtran}
\bibliography{biblio}





\end{document}